# Phase stability of lanthanum orthovanadate at high-pressure

D. Errandonea[1,†], J. Pellicer-Porres[1], D. Martínez-García[1], J. Ruiz-Fuertes[1,2], A. Friedrich[2,††], W. Morgenroth[2], C. Popescu[3], P. Rodríguez-Hernández[4], A. Muñoz[4], M. Bettinelli[5]

[1] Departamento de Física Aplicada-ICMUV, MALTA Consolider Team, Universidad de Valencia, Edificio de Investigación, C/Dr. Moliner 50, Burjassot, 46100 Valencia, Spain

[2] Institut für Geowissenschaften, Goethe-Universität Frankfurt, Altenhöferallee 1, 60438 Frankfurt am Main, Germany

[3] CELLS-ALBA Synchrotron Light Facility, 08290 Cerdanyola, Barcelona, Spain

[4] Departamento de Física, Instituto de Materiales y Nanotecnología, MALTA Consolider Team, Universidad de La Laguna, 38205 La Laguna, Tenerife, Spain

[5] Laboratory of Solid State Chemistry, DB and INSTM, Università di Verona, Strada Le Grazie 15, 37134, Verona, Italy

**Abstract:** When monoclinic monazite-type $LaVO_4$ (space group $P2_1/n$) is squeezed up to ~12 GPa at room temperature, a phase transition to another monoclinic phase has been found. The structure of the high-pressure phase of $LaVO_4$ is indexed with the same space group ($P2_1/n$), but with a larger unit-cell in which the number of atoms is doubled. The transition leads to an 8% increase in the density of $LaVO_4$. The occurrence of such a transition has been determined by x-ray diffraction, Raman spectroscopy, and *ab initio* calculations. The combination of the three techniques allows us to also characterize accurately the pressure evolution of unit-cell parameters and the Raman (and IR)-active phonons of the low- and high-pressure phase. In particular, room-temperature equations of state have been determined. The changes driven by pressure in the crystal structure induce sharp modifications in the color of $LaVO_4$ crystals, suggesting that behind the monoclinic-to-monoclinic transition there are important changes of the electronic properties of $LaVO_4$.

[†] Corresponding author, email: daniel.errandonea@uv.es

[††] Present address: Institut für Anorganische Chemie, Julius-Maximilians-Universität Würzburg, Am Hubland, 97074 Würzburg, Germany

I. Introduction

Rare-earth orthovanadates are important materials for existing and future technologies. Recently, they have attracted considerable attention because of their potential application in alternative green technologies [1]. Orthovanadates are also suitable candidates for laser-host materials [2]. On top of that, they exhibit properties like luminescence, chemical stability, and non-toxicity, which make them, as nanoparticles, promising materials for biomedical applications as controlled drug delivery into the human body [3]. $LaVO_4$ is one of the members of this family of compounds. Typically, it adopts the monazite-type structure (space group $P2_1/n$, Z = 4) shown in Fig. 1 [4]. However, it can also be obtained by the hydrothermal method in the tetragonal zircon-type structure (space group $I4_1/amd$, Z = 4) as a metastable phase at ambient conditions [5].

The study of rare-earth orthovanadates under compression has become the subject of a large interest in the last decade [5 - 28]. Several pressure-induced structural transitions have been discovered, which have consequences on the physical properties of orthovanadates. These consequences include an electronic band-gap collapse [14] and pressure-induced metallization [10]. Among orthovanadates, one of the compounds whose properties have been less explored under compression is $LaVO_4$. Three high-pressure (HP) powder x-ray diffraction (XRD) experiments have been reported in the literature. These experiments were performed using either a 16:3:1 methanol–ethanol–water mixture or a 4:1 methanol-ethanol mixture as pressure-transmitting medium (PTM). In one of them the compressibility of the monazite structure has been explored up to 8 GPa [18]. In the other, pressure was extended up to 12 GPa [9], and possible evidences of an isosymmetric structural phase transition at 8.6 GPa were described. In the third experiment, zircon-type $LaVO_4$ was found to transform to monazite-type $LaVO_4$ at 5 GPa [5]. A post-monazite phase appeared at 12.9 GPa; however, the quality of the XRD patterns obtained under non-hydrostatic conditions did not allow the identification of the crystal structure of the HP phase. Raman spectroscopy measurements have been carried out at ambient conditions for monazite-type LaVO$_4$ [29 – 32]. High-pressure Raman measurements have been recently carried out for zircon-type and monazite-type LaVO$_4$ using a non-hydrostatic PTM (silicone oil) up to 18 GPa [5, 25]. In monazite-type LaVO$_4$ approximately half of the Raman-active modes have been measured, and only some of them have been measured under compression [25]. Additionally, there are many issues on the mode assignment that need to be clarified yet. On top of that, some contradictions can be found among the different experiments. Regarding theoretical studies, only the physical properties of LaVO$_4$ at ambient pressure and the effect of pressure on the unit-cell parameters up to 8 GPa have been studied by means of *ab initio* calculations [18, 33].

The combination of XRD, Raman spectroscopy, and *ab initio* calculations have been shown to be a powerful tool to study the HP behavior of orthovanadates and related oxides [34, 35]. Here we combine the three techniques to study LaVO$_4$ up to 22 GPa. All our experiments were carried out under quasi-hydrostatic conditions. We report the occurrence of the onset of a phase transition at 12 GPa, coexisting the low- and high-pressure phases up to 15.5 GPa. The HP phase remains stable up to 22 GPa. A crystal structure is proposed for the HP phase and the effect of pressure on the vibrational and structural properties of the two phases is discussed. The reported results will contribute to the better understanding of the HP behavior of orthovanadates and related AXO$_4$ ternary oxides.

II. Experimental Details

LaVO$_4$ crystals were prepared by the flux growth method using Pb$_2$V$_2$O$_7$ as the solvent. Pure La$_2$O$_3$, V$_2$O$_5$, PbO and Na$_2$B$_4$O$_7$ were used as starting materials (99.99 % Aldrich). The composition of the growth mixtures was (in molar percent ratio):

La$_2$O$_3$:V$_2$O$_5$:PbO:Na$_2$B$_4$O$_7$ = 2.3:31.5:62.9:3.3. After careful mixing the starting mixtures were put in Pt crucibles and heated to 1270 °C in a horizontal programmable furnace. They were maintained at this temperature for 12 h (soaking time), then cooled to 800 °C at a rate of 1.8 °C/h and finally to room temperature at a rate of 15 °C/min. Transparent crystals in the form of platelets having an average size $0.5 \times 2 \times 2$ mm$^3$ were separated from the flux by dissolving it in hot diluted HNO$_3$. The obtained crystals were confirmed to have a monazite-type structure by powder XRD (PXRD) measurements. Neither minority phases nor impurities were detected in the crystals. The lattice parameters are $a$ = 7.021(2) Å, $b$ = 7.242(2) Å, $c$ = 6.704(2) Å, and $\beta$ = 104.875(5)°, which are in good agreement with the literature [4, 9, 18]. Single-crystal XRD (SXRD) measurements were also used to determine the orientation of the crystals. These SXRD measurements were carried out using a Nonius Kappa CCD diffractometer, employing Mo K$_\alpha$ ($\lambda$ = 0.71073 Å) radiation.

Ambient pressure and HP Raman experiments were performed with LaVO$_4$ single crystals. The sample dimensions were 3 $x$ 3 $x$ 0.5 mm$^3$ for the first set of experiments and 50 $x$ 50 $x$ 10 μm$^3$ for the HP experiments. Raman spectra were excited using a He-Ne laser ($\lambda$ = 632.8 nm). The incident power on the sample was 10 mW. The Raman set-up consists of a confocal microscope, an edge filter, a 1 m focal-length spectrometer equipped with a 600 groves/mm grating (TRH 1000, JobinYvon), and a thermoelectric-cooled multichannel CCD detector. The spectral resolution is below 2 cm$^{-1}$. Wavenumbers were systematically calibrated in the whole spectral range using the laser plasma lines. For the polarized Raman spectroscopy experiments at ambient pressure we used crystals with two different orientations with respect to the incident laser radiation. In the first orientation the incidence direction was chosen along the *b*-axis, which we label *y*. The principal axis perpendicular to the *b*-axis was optically determined placing the sample between two crossed linear polarizers. The polarization labeled *z* corresponds to the sample orientation which did not rotate the polarization of the He-Ne laser. The polarization denoted by *x* is perpendicular to *y* and *z*. In the second orientation the incidence was along the [$10\bar{1}$] direction (labelled *x'*). Now the polarization was set either parallel to *b* (*y*) or in the perpendicular direction *z'*.

Under compression, two independent sets of non-polarized Raman measurements were performed up to 20.8 GPa using the same Raman set up as for the ambient pressure measurements. In one set of experiments the $LaVO_4$ crystal was loaded inside a membrane diamond-anvil cell (DAC) equipped with 350 μm diamond culets. 16:3:1 methanol−ethanol−water was used as PTM and the pressure chamber was a 100 μm hole drilled in a pre-indented (40 μm) Inconel gasket. The second set of experiments was performed using a Boehler-Almax DAC equipped with 350 μm diamond culets. In this case, we employed a powder sample and Ne as PTM and a tungsten gasket with a 100 μm hole (30 μm thickness).

HP SXRD and PXRD experiments were carried out using the Boehler-Almax DAC employed for Raman experiments. For SXRD, $LaVO_4$ crystals were loaded together with a ruby chip and Ne as PTM in 120 μm-diameter holes drilled in a tungsten gasket pre-indented to a thickness of 40 μm. The same procedure was used in the DAC preparation for PXRD measurements. For these experiments, we loaded in the DAC a pellet made from a finely ground powder obtained from single crystals of $LaVO_4$.

SXRD experiments were performed at 3.9 and 11.5 GPa in the low-pressure phase. They were carried out at the Extreme Conditions Beam-line P02.2 at PETRA III. We used a monochromatic x-ray beam ($\lambda$ = 0.28968 Å) focused down to 2.4 (horizontal) × 2.1 (vertical) μm$^2$ (FWHM) by a pair of Kirkpatrick-Baez mirrors. The diffraction images were collected with a Perkin-Elmer XRD 1621 detector located at 399.82 mm from the sample. The measurements were performed using an increment of 1 °/step in ω using the complete opening angle of the DAC (85°). The image format was converted according to the procedure described by Rothkirch *et al*. [36]. The indexing of the Bragg reflections and the intensity data reduction were done with the CrysAlis software [37]. The crystal structure was refined with SHELX97-2 [38] with anisotropic displacement parameters for all the atoms and starting parameters taken from Ref. 4.

PXRD experiments were carried out with the aim of solving the crystal structure of the HP phase of $LaVO_4$ and to study its HP behavior. Therefore, the starting pressure was 11.5 GPa, a pressure at which the low-pressure phase was confirmed by single-crystal experiments. The maximum pressure reached in these experiments was 22 GPa. PXRD experiments were performed at the MSPD beam-line at ALBA synchrotron facility [39]. The beam-line is equipped with Kirkpatrick-Baez mirrors to focus the monochromatic beam to 20 μm × 20 μm and a Rayonix CCD detector with an active area of 165 mm diameter. We used a wavelength of 0.534 Å, and the sample−detector distance was set to 280 mm. The two-dimensional diffraction images were integrated with FIT2D software [40]. Structural analysis was performed with PowderCell [41] and Fullprof [42]. For Rietveld refinements first the background was fitted with a Chebyshev polynomial function of first kind with eight coefficients and then subtracted. In addition, since the occupancy and the atomic displacement factors are correlated and sensitive to background subtraction, they were constrained to 1 and $B = 0.5 \text{ Å}^2$, where B is the overall displacement factor [43]. On top of that, since the number of reflections is not enough for the reliable refinement of all the atomic coordinates and other structural parameters, we assumed for the monazite-type phase the atomic positions obtained from SXRD. For the HP phase the atomic positions of the calculated model structure obtained from our numerical simulations were used and fixed during the refinement.

In all the HP experiments pressure was determined using the ruby fluorescence method [44]. In HP PXRD experiments performed at ALBA we confirmed the pressure determined with the ruby using the Ne Bragg reflections that can be observed after the solidification of Ne and the equation of state of Ne [45].

### III. Ab initio calculations

Total energy *ab initio* simulations have been performed within the framework of density-functional theory (DFT) [46]. We used the Vienna *Ab Initio* Simulations Package, VASP [47], to carry out calculations with the pseudopotential method. The projector augmented wave scheme (PAW) [48] was employed to take into account the full nodal character of the all-electron charge density distribution in the core region. In order to obtain accurate converged results a set of plane waves up to a kinetic energy cutoff of 520 eV was used. The exchange-correlation energy was described in the generalized-gradient approximation (GGA) with the PBE for solids [49] (PBESol) prescription. A dense grid of Monkhorst-Pack [50] *k*-special points was used to perform integrations in the Brillouin zone (BZ) in order to obtain very well converged results for the energies and forces. At a selected volume we fully relaxed the structures through the calculation of the forces on atoms and the stress tensor. In the optimized structures the forces on the atoms are less than 0.005 eV/Å and deviations of the stress tensor from a diagonal hydrostatic form are less than 0.1 GPa. It is important to note that from the simulation we obtain total energy (E) as a function of volume (V), and the corresponding pressure (P), from which we determined the enthalpy (H). The thermodynamically stable phase at different pressures is obtained from the P-H curves of the analyzed candidate crystal structures. This methodology has been applied successfully to the study of several orthovanadates [8, 12, 14, 15, 18, 22, 27, 51].

Phonon calculations were performed at the zone centre (Γ point) of the Brillouin zone (BZ). We used the direct method [52] where we need highly converged results on forces for the calculation of the dynamical matrix. The construction of the dynamical matrix at the Γ point of the BZ involves separate calculations of the forces in which a fixed displacement from the equilibrium configuration of the atoms is considered. The number of such independent displacements is reduced due to the crystal symmetry. Diagonalization of the dynamical matrix provides the frequencies of the normal modes. Moreover, these calculations provide the symmetry and eigenvectors of the vibrational modes in each structure at the Γ point.

**IV. Results and discussion**

**Single-crystal XRD experiments**

Previous PXRD experiments on $LaVO_4$ [5, 9] found that the PXRD patterns can be assigned to the monazite structure up to 11.6 GPa. However, a discontinuity in the unit-cell volume was detected at 8.6 GPa [9]. As a possible explanation of this volume collapse the occurrence of an isomorphic phase transition was proposed. More recent PXRD experiments reported the transition to a post-monazite phase in $LaVO_4$ nanorods [5]. However, the crystal structure of this HP phase remains unsolved. SXRD experiments are more accurate than powder XRD experiments to determine crystal structures. We performed SXRD at 3.9 and 11.5 GPa to compare the crystal structures of the two isomorphic phases previously reported [9]. However, our experiments showed that the crystal structure of $LaVO_4$ remains in the low-pressure monazite-type at both pressures, contradicting previous PXRD experiments [9]. The lattice parameters and the atomic positions extracted from the HP SXRD experiments are provided in Table 1. They are compared with those we obtained from PXRD at ambient pressure and with calculations. Fig. 2 also shows the unit-cell parameters at different pressures obtained from both PXRD and SXRD experiments. We found that unit-cell parameters and atomic coordinates gradually change with pressure. In particular, the unit-cell parameters determined at 3.9 GPa from SXRD agree very well with those previously determined from PXRD experiments [9, 18]. The agreement with calculations is also good between experiments and calculations (Table 1). The unit-cell parameters determined at 11.5 GPa agree with our theoretical calculations showing a gradual change with pressure and not showing the discontinuity previously reported [9]. This fact suggests that the previously proposed isostructural transition at 8.6 GPa [9] is not of intrinsic origin but likely originates from stress between grains in the HP PXRD experiments [53, 54], which in turn underlines the importance of the SXRD experiments.

**Raman spectroscopy at ambient pressure**

Ambient-pressure polarized Raman spectra of LaVO$_4$ single crystals are shown in Fig. 3. According to group theory, monazite-type LaVO$_4$ has 72 vibrational modes. All the atoms occupy general 4e Wyckoff position. Using the correlation method, the symmetry decomposition of zone center phonons is then dictated by the factor group 2/*m*:

$$\Gamma = 18\,A_g + 18B_g + 18A_u + 18B_u$$

Three modes ($A_u + 2B_u$) correspond to acoustic vibrations. There are 33 infrared-active modes ($17A_u + 16B_u$) and 36 Raman-active modes ($18A_g + 18B_g$). In this work we consider the unique axis as the *b*-axis. The Raman tensors are then:

$$A_g = \begin{pmatrix} b & 0 & d \\ 0 & c & 0 \\ d & 0 & a \end{pmatrix} \quad B_g = \begin{pmatrix} 0 & f & 0 \\ f & 0 & e \\ 0 & e & 0 \end{pmatrix}$$

The Raman spectra corresponding to different backscattering configurations are plotted in Fig. 3. We have used selection rules to identify the symmetry of modes. In all the configurations used, except for x'(y,z')x' and x'(z',y)x', $A_g$ are allowed and $B_g$ are forbidden. In the x'(y,z')x' and x'(z',y)x' scattering geometries, the situation is inversed, being the $B_g$ modes the only ones which are allowed. In Fig. 3, the $B_g$ modes are observed in the spectra labelled x'(y,z')x' and x'(z',y)x'. There are traces of the most intense $A_g$ peaks where we expect only $B_g$ modes. This is evident with the most intense $A_g$ peak at 859 cm$^{-1}$. The polarization leakage is related to the focussing objectives used in our microscopic system, which establish a propagation direction not exactly parallel to the optical axis. There are two weak $B_g$ modes in the spectra labelled x′(y,z′)$\bar{x}$′ and x′(z′,y)$\bar{x}$′, at 440 and 855 cm$^{-1}$, which are observed as shoulders of more intense $A_g$ peaks. Finally, the $A_g$ mode at 102 cm$^{-1}$ is so weak that its identification should be considered as tentative. All the Raman modes measured here are given in Table 2.

There have been a few works [25, 29 - 32] on the Raman spectra of monazite-type LaVO$_4$. Good agreement is observed between our data and those of Refs. 29, 31, and 32. A comparison of our frequencies with those by Jia *et al.* [29], who reported the non-polarized Raman spectra of Eu-doped LaVO$_4$ nanocrystals, is shown in Table 2. Most of the modes given in Ref. 29 correspond to $A_g$ modes. In contrast, the results reported in Ref. 25 (Table 2) show Raman frequencies that differ up to 10 cm$^{-1}$ from those reported by us and other authors [29, 31, 32]. The Raman spectrum of Er-doped LaVO$_4$ seems to be quite different at all [30]. In particular, in Er-doped LaVO$_4$ the most intense mode is located at 912 cm$^{-1}$, clearly higher than in our work (859 cm$^{-1}$) or in the rest of the literature (861 cm$^{-1}$ to 855 cm$^{-1}$) [29, 31, 32]. This seems to be attributed to the substitution of La by Er. The highest phonon in tetragonal ErVO$_4$, however, it is situated at 890 cm$^{-1}$ [55] well below 900 cm$^{-1}$. If considering a hypothetical local formation of ErVO$_4$, the highest frequency phonon in the monoclinic monazite-type phase is expected at a wavenumber even smaller than that of tetragonal ErVO$_4$, as was found in PrVO$_4$ [16]. All these points raise doubts about the influence of substitutional Er in the Raman spectrum of LaVO$_4$.

The strong V-O bond in the VO$_4^{-3}$ tetrahedron has provided the depart point in the usual classification of the vibrational spectrum of orthovanadates [56]. The normal modes of the VO$_4^{-3}$ tetrahedron have been measured [57] in an aqueous solution of Na$_3$VO$_4$. They are $\nu_1$(A$_1$) = 870 cm$^{-1}$, $\nu_2$(E) = 345 cm$^{-1}$, $\nu_3$(F$_2$) = 825 cm$^{-1}$, and $\nu_4$(F$_2$) = 480 cm$^{-1}$. The symbols in parentheses denote the irreducible representations of the T$_d$ group. In LaVO$_4$ the tetrahedral symmetry is lost and most of the modes display complex vibration patterns. However, there are some modes which are clearly related to the tetrahedral symmetry; as shown by our *ab initio* calculations (Table 2). This is the case for the most intense A$_g$ mode at 859 cm$^{-1}$, which is related to the $\nu_1$ breathing mode. On the same line, in the intense A$_g$ mode at 374 cm$^{-1}$, the opposed O-V-O bonds in the VO$_4$ tetrahedron bend in phase, as in the $\nu_2$ vibrations. Additionally, the 440 cm$^{-1}$ mode retains the $\nu_4$ essence, as the O atoms oscillate along the tetrahedron edges in this mode. An additional comment to make is that there is not any mode showing a pure $\nu_3$ characteristic. There is always a mixing between the $\nu_1$ and $\nu_3$ features.

The vibrations in LaVO$_4$ can also be classified in terms of stretching modes, bending modes, and rotational-translational modes of the VO$_4$ tetrahedron. We observe an isolated group of phonons in the wavelength range from 768 to 862 cm$^{-1}$ which correspond to stretching modes. As an example, the *ab initio* calculations show that the highest mode corresponds basically to the stretching of the shortest V-O bond. The bending modes are observed in the 309-440 cm$^{-1}$ range. A typical example is the above mentioned A$_g$ mode at 374 cm$^{-1}$. To end, the rotational-translational lattice modes display wavenumbers from 64 to 252 cm$^{-1}$. Here, the most remarkable fact is the clear increase in the La vibrational amplitude as the wavenumber of the phonon decreases.

**Raman spectroscopy at high-pressure**

Out of the 36 Raman modes observed in monazite-type LaVO$_4$ when the sample was loaded in the DAC, only the strongest 17 modes could be followed under compression. Fig. 4 shows a selection of non-polarized Raman spectra measured during the upstroke. The spectra measured from ambient pressure up to 11.2 GPa are qualitatively similar to the Raman spectra measured at ambient conditions. All of them can be assigned to the monazite-type phase of LaVO$_4$. The pressure evolution of these modes is represented in Fig. 5. The frequencies ($\omega$) and pressure coefficients (d$\omega$/dP) of the different modes are summarized in Table 2, where they are compared with our calculations and with the experimental results obtained under non-hydrostatic conditions [25]. As expected, calculations show a better agreement with our quasi-hydrostatic experiments than with previous non-hydrostatic experiments. In the experiments we have found that most modes harden under compression except for one of the low-frequency $A_g$ modes which has a negative pressure coefficient. The highest pressure coefficients are some of the internal modes of the $VO_4$ tetrahedron (see Table 2). In general, the pressure behavior of Raman modes of monazite-type $LaVO_4$ is qualitatively similar to that of isomorphic $PbCrO_4$ [58] and $CePO_4$ [59]. In particular, in all of them the internal stretching modes have pressure coefficients of the same order. In addition, in lead chromate there is also a lattice mode that slightly softens under compression as found by us in $LaVO_4$. To close this part of the discussion, we would like to add that not surprisingly the internal modes of the $VO_4$ tetrahedron have similar frequencies and pressure-coefficients in monazite-type and zircon-type $LaVO_4$ [24].

Changes in the Raman spectrum, including the appearance of additional peaks are clearly detected at 12.2 GPa, suggesting the onset of a phase transition (Fig. 4). The main changes are the increase of the number of Raman modes, the small drop in the wavenumber of the highest frequency mode and the appearance of extra bands from 500 cm$^{-1}$ to 700 cm$^{-1}$ (Fig. 5) in the phonon gap of the low-pressure phase. Changes with pressure are gradual since both phases coexist from 12 GPa to 14.2 GPa, with the HP phase appearing as a single phase at 15.4 GPa. There is no evidence of a second pressure-induced phase transition up to 20.8 GPa. Upon decompression the changes are reversible but a large hysteresis is detected. In particular, the bands of the HP phase can be observed down to 5 GPa upon decompression. At ambient pressure, the Raman spectrum of the monazite-type phase is recovered as can be seen in Fig. 4. Consequently, the observed phase transition is reversible. From our experiments we identified 31 Raman phonons of the HP phase and their pressure dependence. Their frequencies and pressure coefficients are summarized in Table 3. Due to the large number of modes and their broadening under compression, we are cautious about the mode assignments and accuracy of the pressure coefficients reported in the table. In comparison with the low-pressure phase, the pressure coefficients of the high-pressure phase tend to be larger. In addition, they are all positive in our experiment. In the high-frequency region there are seven modes, which are the most intense modes and could be probably related to the internal modes of the coordination polyhedron of the V atom. The fact that these modes are distributed in a larger frequency region than the internal stretching modes of the monazite-type phase is related to a coordination increase of V (see next section). The possible mode assignment will be discussed when presenting the results of *ab initio* calculations.

When measuring the Raman spectra we have observed color changes in the $LaVO_4$ single crystal at the onset of the phase transition and at higher pressures. Up to 11.2 GPa, the crystal of $LaVO_4$ is colorless and transparent, which is consistent with the fact that $LaVO_4$ has a band gap of 3.5 eV [14]. At 12.2 GPa, after the transition onset, the sample becomes yellowish, which indicates that the phase transition induces a band gap collapse, which is consistent with the HP behavior of the band gap in other monazite-type oxides [60]. Finally, as pressure keeps increasing, $LaVO_4$ becomes orange, suggesting a red-shift of the band gap with pressure in the HP phase. Clearly, the observed phase transition involves important changes of the electronic properties of $LaVO_4$, which might be interesting issues for future studies.

**Powder XRD experiments: The crystal structure of the high-pressure phase**

In contrast to previous powder-XRD experiments we did not find evidences of structural changes in $LaVO_4$ up to 12.2 GPa in both our present single-crystal XRD experiments and Raman experiments. Therefore, in order to better determine the transition pressure and the crystal structure of the HP phase, we have carried out PXRD experiments at higher pressures. Fig. 6 shows the results of the Rietveld refinement of the measured PXRD pattern at 11.5 GPa for the monazite-type phase of LaVO$_4$ and at 16 GPa for the high-pressure phase. Also Bragg peaks of Ne are detected due to its solidification. The goodness-of-fit parameters obtained from the refinement are R$_p$ = 3.92 %, R$_{wp}$ = 4.49 % and the reduced $\chi^2$ = 1.25 at 11.5 GPa. The unit-cell parameters of monazite-type LaVO$_4$ at 11.5 GPa are $a$ = 6.786(2) Å, $b$ = 7.041(3) Å, $c$ = 6.545 (2) Å, and β = 103.75°, which agree well with the results of the SXRD experiments (Table 1).

At 12 GPa we detected the appearance of extra weak reflections, which suggest the onset of a phase transition, in good agreement with Raman experiments. The coexistence of the low- and high-pressure phases was observed up to 15.5 GPa, while from 16 to 22 GPa only the HP phase was observed. Fig. 6 illustrates the differences between the diffraction patterns of the low- and high-pressure phases. In particular, the appearance of Bragg reflections at low angles suggests an increase of the size of the unit-cell (Fig. 6). The DICVOL routine included in Fullprof [42] was used to index the Bragg reflections of the HP phase of LaVO$_4$ found below 2θ = 15° to avoid the overlapping of LaVO$_4$ and Ne reflections. The highest figure of merit was obtained for a monoclinic unit cell with $a$ = 12.288(9) Å, $b$ = 6.491(5) Å, $c$ = 6.836(5) Å, and $β$ = 95.62(9)°. The analysis of the systematic absences (0k0, k = 2n and h0l, h + l = 2n) indicated $P2_1/n$ as the possible space group for the HP phase. While this space group is the same as for the low-pressure phase, the unit-cell volume is doubled in the high-pressure phase. The same space group is observed in the AgMnO$_4$-type structure, which is known for being a HP post-monazite structure in CaSO$_4$ and CaSeO$_4$ [61, 62], and in the BaWO$_4$-II structure, which has been found in BaWO$_4$ and related oxides [43]. In our case, the determined unit-cell parameters resemble those of the BaWO$_4$-II structure. A LeBail fit confirmed that the proposed unit-cell and space group are sufficient to reproduce the measured PXRD pattern ($R_P$ = 4.12 %, $R_{WP}$ = 4.75 %, $\chi^2$ = 1.36). Our *ab initio* calculations support the assignment of the $BaWO_4$-II -type structure as the crystal structure of the HP phase of $LaVO_4$, and also show that the $BaWO_4$-II -type phase is the most stable phase of $LaVO_4$ beyond 10.5 GPa (see next section). A Rietveld refinement resulted in small residuals (Fig. 6). The refined unit-cell parameters are $a$ = 12.289(9) Å, $b$ = 6.492(5) Å, $c$ = 6.836(5) Å, and $\beta$ = 95.6(1)°. The goodness-of-fit parameters are $R_P$ = 4.98 %, $R_{WP}$ = 5.97 %, $\chi^2$ = 1.65. In summary, the Rietveld refinements of the XRD patterns measured for both phases clearly demonstrate the occurrence of a monoclinic-to-monoclinic transition.

To conclude this section we will comment on the compressibility of the low- and high-pressure phase. In Fig. 2 we show the unit-cell parameters obtained at different pressures (open symbols are from this work). It can be seen that for the low-pressure phase our results agree well with those obtained in Refs. 9 and 18. In addition, our study shows that up to 15.5 GPa the unit-cell parameters change gradually, the compression being slightly anisotropic. We also obtain that the unit-cell volumes we measure here at 11.5, 12, and 15.5 GPa agree well will the extrapolation of the previously determined EOS [9, 18]. Regarding the HP phase, we found that the phase transition implies a collapse of around 8 % for the normalized volume (taking into account Z = 4 in the low-pressure phase and Z = 8 in the HP phase). The volume difference between the two phases, when they coexist at the same pressure, indicates a first-order phase transition. In particular, according with our structural refinements and calculations the phase transition implies an increase of the coordination number of V from [4] to [5+1] and of the coordination number of La from [9] to [10] (see Fig. 1). The transition also involves a drastic change of the unit-cell parameters as can be seen in Figs. 1 and 2. Based upon the above described facts, we can conclude that the transition involves substantial atomic rearrangements. This means that $LaVO_4$ undergoes a reconstructive transition in spite of the fact that the parent and daughter phases share the same space group. Regarding compressibility, the HP phase appears to be less compressible than the low-pressure phase (see Fig. 2). However, we do not have enough data points to determine properly the EOS of the HP phase. Finally, the compression of the HP phase is nearly isotropic, being the β angle only slightly affected by compression.

**Theoretical results**

We compare now the experimental data presented in the previous sections with the results from our *ab initio* calculations. Figure 7 shows the pressure dependence of the enthalpy difference between the two candidate HP phases of LaVO$_4$ (AgMnO$_4$-type, and BaWO$_4$-II-type) and the low-pressure monazite-type phase. Other structures like zircon-type, scheelite-type, barite-type, and other that appear as HP phases in related compounds [63] have been considered, but they are not energetically competitive with the three structures represented in Fig. 7. This figure shows the monazite-type phase as being stable at zero and low pressure, with $V_0 = 328.2$ Å$^3$, $B_0 = 105.2$ GPa, and $B_0' = 4.3$. The structural parameters of the equilibrium structure at ambient pressure are given in Table 1. The agreement with the experiments is very good for the structure and the P-V equation of state (See Table 1 and Refs. 9 and 18). As pressure increases, the monazite-type structure becomes unstable against both AgMnO$_4$-type and BaWO$_4$-II-type. However, the second structure has always a lower enthalpy than the first one. This structure, which has been also found as a HP phase in BaWO$_4$ [43] and SrMoO$_4$ [64], only emerges as a structurally different and thermodynamically stable phase above a compression threshold of about 10.5 GPa. This is consistent with the phase transition found in the experiments. Structural information on the HP phase is given in Table 4. The calculated unit-cell parameters agree well with those determined from our experiments. From the calculations we also obtained the EOS of the HP phase being $V_0 = 609.2$ Å$^3$, $B_0 = 154$ GPa, and $B_0' = 4.2$. The transition involves a density increase of approximately 8 %. The high-pressure phase has a larger bulk modulus than the low-pressure phase, which is consistent with our experimental observations.

We have also calculated the Raman-active phonons for the low- and high-pressure structure. Results are shown in Tables 2 and 3 and compared with experiments and previous calculations. The mismatch between the calculated and observed frequencies of the Raman modes is below 7% in both low-pressure and HP phases. The HP phase has considerably more phonons than the low-pressure phase as a consequence of the doubling of the unit-cell. For the BaWO$_4$-II type structure, the vibrational modes have the following mechanical representation $\Gamma = 36A_g + 36A_u + 36B_g + 36B_u$, with 72 Raman-active (g) modes, 69 IR-active (u) modes, and three acoustic modes (1 $A_u$ + 2 $B_u$). Out of the 72 Raman modes, in the experiments we observed only 31 modes (approximately the same proportion of modes observed under HP for the low-pressure phase). A similar number of Raman modes have been measured for the same phase in other compounds [43, 64]. Possible reasons for observing fewer modes than expected could be the broadening and overlaping of Raman modes and the presence of low-intensity modes which go below the noise threshold and are consequently not visible. The experimental assignment of the modes symmetry in the BaWO$_4$-II phase is difficult because of the lack of information on polarization inside the DAC, and because the number of modes that can be clearly resolved in the Raman spectra is around 31. We have made a tentative analysis by matching theoretical and experimental frequencies (see Table 3).

Regarding the pressure coefficients, they agree well for the low-pressure phase, while for the HP phase the agreement is not so good. A possible reason for it is the fact that due to the broadening of the Raman bands the accuracy of the experimental determination of the pressure coefficients of some modes may be affected. As can be observed in Table 3, according to theory two modes have negative pressure coefficients.

Unfortunately these modes are not observed in the experiments and therefore this conclusion cannot be confirmed by our measurements.

From the calculations we also obtained the IR-active modes (Tables 5 and 6), which are reported for completeness. In both phases the IR modes show a similar frequency distribution to the Raman modes. In particular, the modes calculated for the low-pressure phase agree rather well with those calculated by Sun *et al.* [33]. In addition, three of our stretching modes agree within 1.5% with the three modes experimentally measured. The experimental wavenumbers are 835, 850, and 881 cm$^{-1}$ and the calculated wavenumbers are 837, 861, and 869 cm$^{-1}$, which correspond to $A_u$, $B_u$, and $A_u$ modes respectively. Calculations also show that IR and Raman modes have comparable pressure coefficients. Several modes existing in both the low- and high-pressure phases show a weak softening under compression (Table 5 and 6).

## V. Concluding Remarks

We reported a combined experimental and theoretical study of the high-pressure structural and vibrational properties of LaVO$_4$. Polarized single-crystal Raman experiments were carried out for the first time identifying the 36 Raman-active modes of the low-pressure monazite phase. HP Raman measurements provide evidence that the previously reported phase transition (8.6 GPa) does not occur up to 12.2 GPa. They also provide information on the pressure dependence of the Raman modes of the low- and high-pressure phases. Single-crystal XRD measurements confirm that the low-pressure phase is stable up to 11.5 GPa and provide information on the evolution of the unit-cell parameters and atomic positions under compression. A pressure-induced phase transition is found to occur at 12 GPa by powder XRD measurements, the HP phase remaining stable up to 22 GPa. From these experiments we solved the crystal structure of the HP phase and obtained the pressure dependence of its lattice parameters. This phase can be described with the same monoclinic space group as the low-pressure phase, but with doubled unit-cell volume. It is isomorphic to BaWO$_4$–II. *Ab initio* calculations are in full agreement with the experiments. They have been an extremely good help for Raman mode assignment and for the identification of the crystal structure of the HP phase. They also provide information on IR-active modes for the low- and high-pressure phases as well as on their pressure dependences. No experimental results are available for these modes yet, therefore our calculations can be a good guide for mode identification in future experiments. Additionally, calculations provide the pressure evolution of lattice parameters and the P-V EOS for the low- and high-pressure phase. These results describe quite well the experimental results from this and previous works. Finally, in our experiments we observed a sharp color change in LaVO$_4$ at the monoclinic-to-monoclinic phase transition. This suggests that important changes of the electronic structure of LaVO$_4$ are associated to the structural phase transition. The changes induced in the electronic properties might be an interesting issue for future studies.

## ACKNOWLEDGEMENTS

This work has been done under financial support from Spanish MINECO under projects MAT2013-46649-C4-1/3-P and MAT2015-71070-REDC. Supercomputer time has been provided by the Red Española de Supercomputación (RES) and the MALTA cluster. The authors thank the SCSIE from Universitat de Valencia the technical support in ambient pressure XRD measurements and ALBA and PETRA III synchrotrons for providing beam-time for the XRD high-pressure experiments. DESY Photon Science is gratefully acknowledged. PETRA III at DESY is a member of the Helmholtz Association (HGF). J. R.-F. thanks the Alexander von Humboldt Foundation for a postdoctoral fellowship and the Spanish MINECO for the support through the Juan de la Cierva program (IJCI-2014-20513). W. M. thanks the support by BMBF through project 05K13RF1.

**References**

[1] Y. Zhang, G. Li, X. Yang, H. Yang, Z. Lu, R. Chen, J. Alloys Compd. **551**, 544 (2013).

[2] S. Tang, M. Huang, J. Wang, F. Yu, G. Shang, J. Wu, J. Alloys Compd. **513**, 474 (2012).

[3] Y. Liang, P. Chui, X. Sun, Y. Zhao, F. Cheng, Ka. Sun, J. Alloys Compd. **552**, 289 (2013).

[4] C.E. Rice, W.R. Robinson, Acta Crystallogr. B **32**, 2232 (1976).

[5] H. Yuan, K. Wang, C. Wang, B. Zhou, K. Yang, J. Liu, and B. Zou, J. Phys. Chem. C **119**, 8364 (2015).

[6] D. Errandonea, R. Lacomba-Perales, J. Ruiz-Fuertes, A. Segura, S. N. Achary, and A. K. Tyagi, Phys. Rev. B **79** 84104 (2009).

[7] D. Errandonea, O. Gomis, B. Garcia-Domene, J. Pellicer-Porres, V. Katari, S. N. Achary, A. K. Tyagi, and C. Popescu, Inorg. Chem. **52**, 12709 (2013).

[8] W. Paszkowicz, O. Ermakova, J. López-Solano, A. Mujica, A. Muñoz, R. Minikayev, C. Lathe, S. Gierlotka, I. Nikolaenko, and H. Dabkowska, J. Phys.: Condens. Matter **26**, 025401 (2014).

[9] D. Errandonea, C. Popescu, S. N. Achary, A. K.Tyagi, and M. Bettinelli, Mater. Res. Bull. **50**, 279 (2014).

[10] A. B. Garg, K. V. Shanavas, B. N. Wani, and S. M. Sharma, J. Solid State Chem. **203**, 273 (2013).

[11] D. Errandonea, R. S. Kumar, S. N. Achary, and A. K. Tyagi, Phys. Rev. B **84**, 214121 (2011).

[12] A. B. Garg, D. Errandonea, P. Rodriguez-Hernandez, S. Lopez-Moreno, A. Muñoz, and C. Popescu, J. Phys.: Condens. Matter **26**, 265402 (2014).

[13] W. Paszkowicz, P. Piszora, Y. Cerenius, S. Carlson, B. Bojanowski, and H. Dabkowska, Synchrotron Radiat. Nat. Sci. **1-2**, 137 (2010).

[14] V. Panchal, D. Errandonea, A. Segura, P. Rodriguez-Hernandez, A. Muñoz, S. Lopez-Moreno, and M. Bettinelli, J. Appl. Phys. **110**, 043723 (2011).

[15] D. Errandonea, F. J. Manjon, A. Muñoz, P. Rodriguez-Hernandez, V. Panchal, S. N. Achary, and A. K.Tyagi, J. Alloys Compd. **577**, 327 (2013).

[16] D. Errandonea, S. N. Achary, J. Pellicer-Porres, and A. K. Tyagi, Inorg. Chem. **52**, 5464 (2013).

[17] A. B. Garg and D. Errandonea, J. Solid State Chem. **226**, 147 (2015).

[18] O. Ermakova, J. Lopez-Solano, R. Minikayev, S. Carlson, A. Kaminska, M. Glowacki, M. Berkowski, A. Mujica, A. Muñoz, and W. Paszkowicz, Acta Crystallogr. B **70**, 533 (2014).

[19] Z. C. Huang, L. Zhang, and W. Pan, J. Solid State Chem. **205**, 97 (2013).

[20] R. Rao, A. B. Garg, and B. N. Wani, J. Phys. Conf. Series **377**, 012010 (2012).

[21] N. N. Patel, A. B. Garg, S. Meenakshi, B. N. Wani, and S. M. Sharma, AIP Conf. Proc. **1349**, 99 (2011).

[22] J. Lopez-Solano, P. Rodriguez-Hernandez, and A. Muñoz, High Pres. Res. **29**, 582 (2009).

[23] N. N. Patel, A. B. Garg, S. Meenakshi, K. K. Pandey, B. N. Wani, and S. M. Sharma, AIP Conf. Proc. **1313**, 281 (2010).

[24] V. Panchal, D. Errandonea, F.J. Manjon, A. Muñoz, P. Rodriguez-Hernandez, M. Bettinelli, S .N. Achary, and A. K. Tyagi, AIP Conf. Proc. **1665**, 030006 (2015).

[25] X. Cheng, D. Guo, S. Feng, K. Yang, Y. Wang, Y. Ren, amd Y. Son, Opt. Mater. **49**, 32 (2015).

[26] A. B. Garg and D. Errandonea, J. Solid State Chem. **226**, 147 (2015).

[27] C. Popescu, A. B. Garg, D. Errandonea, J. A. Sans, P. Rodriguez-Hernández, S. Radescu, A. Muñoz, S. N. Achary, and A. K. Tyagi, J. Phys.: Condens. Matter **28**, 035402 (2016).

[28] W. Paszkowicz, J. Lopez-Solano, P. Piszora, B. Bojanowski, A. Mujica, A. Muñoz, Y. Cerenius, S. Carlson, and H. Dabkowska, J. Alloys Compd. **648**, 1005 (2015).

[29] C.-J. Jia, L.-D. Sun, Z.-G. Yan, Y.-C. Pang, S.-Z. L, and C.-H. Yan, Eur. J. Inorg. Chem. **2010**, 2626 (2010).

[30] R. Lisiecki, W. Ryba-Romanowski, E. Cavalli, and M. Bettinelli, J. Lumin. **130**, 131 (2010).

[31] B. Xie, G. Lu, Y. Wang, Y. Guo, and Y. Guo, J. Alloys Compd. **544**, 173 (2012).

[32] P. Gangwar, M. Pandey, S. Sivakumar, R. G. S. Pala, and G. Parthasarathy, Cryst. Growth Design **13**, 2344 (2013).

[33] L. Sun, X. Zhao, Y. Li, P. Li, H. Sun, X. Cheng, and W. Fan, J. Appl. Phys. **108**, 093519 (2010).

[34] D. Errandonea, Cryst. Res. Techn. **50**, 729 (2015).

[35] D. Errandonea, C. Popescu, A. B. Garg, P. Botella, D. Martinez-García, J. Pellicer-Porres, P. Rodríguez-Hernández, A. Muñoz, V. Cuenca-Gotor, and J. A. Sans, Phys. Rev. B **93** (2016).

[36] A. Rothkirch, G. D. Gatta, M. Meyer, S. Merkel, M. Merlini, and H.-P. Liermann, J. Synchrotron Rad. **20**, 711 (2013).

[37] Agilent, CrysalisPro software system, version 1.171.36.28, Agilent Technologies UK Ltd., Oxford, UK (2013).

[38] G. M. Sheldrick, Acta Crystallogr. A **64**, 112 (2008).

[39] F. Fauth, I. Peral, C. Popescu, and M. Knapp, Powder Diffraction **28**, S360 (2013).

[40] A P Hammersley, S O Svensson, M Hanfland, A N Fitch, and D Häusermann, High Pressure Research **14**, 235 (1996).

[41] W. Kraus and G. Nolze, J. Appl. Cryst. **29**, 301 (1996).

[42] J. Rodriguez-Carvajal, Physica B **192**, 1 (1993).

[43] O. Gomis, J. A. Sans, R. Lacomba-Perales, D. Errandonea, Y. Meng, J. C. Chervin, and A. Polian, Phys. Rev. B **86**, 054121 (2012).

[44] H. K. Mao, J. Xu, and P. M. Bell, J. Geophys. Res. **91**, 4673 (1986).

[45] A. Dewaele, F. Datchi, P. Loubeyre, and M. Mezouar, Phys. Rev. B **77**, 094106 (2008).

[46] P. Hohenberg and W. Kohn. Phys. Rev. **136**, 3864 (1964).

[47] G. Kresse and J. Hafner, Phys. Rev. B **47**, 558 (1993); G. Kresse and J. Furthmüller, J. Comput. Mater. Sci. **6**, 15 (1996).

[48] P. E. Blöchl. Phys. Rev. B. **50**, 17953 (1994); G. Kresse and D. Joubert, Phys. Rev. B **59**, 1758 (1999).

[49] J. P. Perdew, A. Ruzsinszky, G. I. Csonka, O. A. Vydrow, G. E. Scuseria, L. A. Constantin, X. Zhou, and K. Burke. Phys. Rev. Lett. **100**, 136406 (2008).

[50] H. J. Monkhorst and J. D. Pack, Phys. Rev. B **13**, 5188 (1976).

[51] Y. Z. Chen, S. Li, L. Z. W. Men, Z. L. C. L. Sun, Z. W. Li, and M. Zhou, Acta Phys. Sinica **62**, 246101 (2013).

[52] K. Parlinski, Computer Code PHONON. See: http://wolf.ifj.edu.pl/phonon.

[53] D. Errandonea, A. Muñoz, and J. Gonzalez-Platas, J. Appl. Phys. **115**, 216101 (2014).

[54] M. Guennou, P. Bouvier, P. Toulemonde, C. Darie, C. Goulon, P. Bordet, M. Hanfland, and J. Kreisel, Phys. Rev. Lett. **112**, 075501 (2014).

[55] I. Guedes, Y. Hirano, M. Grimsditch, N. Wakabayashi, C.-K. Loong, and L. A. Boatner, J. Appl. Phys. **90**, 1843 (2001).

[56] S. A. Miller, H. H. Caspers, and H. E. Rast, Phys. Rev. **168**, 964 (1968).

[57] H. Siebert, Zeitschrift für Anorganische und Allgemeine Chemie **275**, 210 (1954).

[58] E. Bandiello, D. Errandonea, D. Martinez-Garcia, D. Santamaria-Perez, and F. J. Manjon, Phys. Rev. B **85**, 024108 (2012).

[59] T. Huang, J.-S. Lee, J. Kung, and C.-M. Lin, Solid State Commun. **150**, 1845 (2010).

[60] D. Errandonea, E. Bandiello, A. Segura, J.J. Hamlin, M.B. Maple, P. Rodriguez-Hernandez, A. Muñoz, J. Alloys Compd. **587**, 14 (2014).

[61] S. López-Moreno, D. Errandonea, P. Rodríguez-Hernández, and A. Muñoz, Inorg. Chem. **54**, 54 (2015).

[62] W. A. Crichton, J. B. Parise, S. M. Antao, and A. Grzechnik, A. Mineral. Mag. **90**, 27 (2005).

[63] D. Errandonea and F.J. Manjon, Progress in Materials Science **53**, 711 (2008).

[64] D. Errandonea, L. Gracia, R. Lacomba-Perales, A. Polian, and J. C. Chervin, J. Appl. Phys. **113**, 123510 (2013).

**Table 1.** Crystal structure of monazite-type LaVO$_4$ (space group $P2_1/n$) at different pressures as obtained from experiments (top) and calculations (bottom).

| | Experiments | | |
|---|---|---|---|
| Pressure | Ambient (powder XRD) | 3.9(1) GPa (single-cystal XRD) | 11.5(1) GPa (single-cystal XRD) |
| $a$ | 7.044(2) Å | 6.9410(9) Å | 6.7890(9) Å |
| $b$ | 7.283(2) Å | 7.1951(9) Å | 7.0406(9) Å |
| $c$ | 6.724(2) Å | 6.6591(9) Å | 6.5448(9) Å |
| $\beta$ | 104.86(5)° | 104.70(5)° | 103.72(5)° |
| La (4e) | (0.2771(4), 0.1570(2), 0.1038(2)) | (0.2743(2), 0.1588(1), 0.1076(1)) | (0.2713(2), 0.1609(1), 0.1141(1)) |
| V (4e) | (0.2995(4), 0.1648(2), 0.6145(6)) | (0.2997(2), 0.1665(1), 0.6173(3)) | (0.2973(2), 0.1688(1), 0.6219(3)) |
| O$_1$ (4e) | (0.2425(5), -0.0009(5), 0.4253(8) | (0.2423(2), -0.0009(1), 0.4287(2)) | (0.2313(2), 0.0001(1), 0.4307(2)) |
| O$_2$ (4e) | (0.3866(7), 0.3434(7), 0.4944(8)) | (0.3854(2), 0.3454(2), 0.4951(2)) | (0.3856(2), 0.3486(2), 0.4954(2)) |
| O$_3$ (4e) | (0.4815(9), 0.1056(5), 0.8247(9)) | (0.4863(2), 0.1093(1), 0.8261(3)) | (0.4898(2), 0.1155(1), 0.8275(3)) |
| O$_4$ (4e) | (0.1190(5), 0.2206(6), 0.7305(9)) | (0.1180(1), 0.2252(2), 0.7338(3)) | (0.1165(1), 0.2314(1), 0.7440(3)) |
| | Calculations | | |
| Pressure | Ambient | 3.89 GPa | 11.71 GPa |
| $a$ | 7.0401 Å | 6.9392 Å | 6.7688 Å |
| $b$ | 7.2774 Å | 7.1931 Å | 7.0432 Å |
| $c$ | 6.6890 Å | 6.6286 Å | 6.5245 Å |
| $\beta$ | 105.014 | 104.723° | 103.851° |
| La (4e) | (0.27764, 0.15702, 0.10405) | (0.27554, 0.15893, 0.10841) | (0.27168, 0.16229, 0.11588) |
| V (4e) | (0.30012, 0.16594, 0.6159) | (0.30046, 0.16820, 0.61779) | (0.29766, 0.17136, 0.62219) |
| O$_1$ (4e) | (0.24415, -0.00218, 0.42623) | (0.23938, -0.00017, 0.42846) | (0.22528, 0.00061, 0.43145) |
| O$_2$ (4e) | (0.38660, 0.34492, 0.49444) | (0.38679, 0.34708, 0.49345) | (0.38589, 0.35046, 0.49237) |
| O$_3$ (4e) | (0.48404), 0.10651, 0.82689) | (0.48851, 0.11077, 0.82818) | (0.49449, 0.11609, 0.82851) |
| O$_4$ (4e) | (0.11791, 0.22261, 0.73008) | (0.11748, 0.22688, 0.73612) | (0.11709, 0.23723, 0.74787) |

**Table 2:** Raman frequencies (ω) and pressure coefficients (dω/dP) for the low-pressure monazite-type structure determined from theory (theo) and experiments (exp). Previous results are included for comparison [[a]Ref. 33, [b]Ref. 29, and [c]Ref. 25]. The Grüneisen parameter was calculated according to $\gamma = \frac{B_0}{\omega_0}\frac{\partial \omega}{\partial P}$ using $B_0 = 105$ GPa.

| Raman mode | $\omega_{theo}$ (cm$^{-1}$) | $d\omega/dP_{theo}$ (cm$^{-1}$/GPa) | $\gamma$ | $\omega_{theo}^{a}$ (cm$^{-1}$) | $\omega_{exp}$ (cm$^{-1}$) | $d\omega/dP_{exp}$ (cm$^{-1}$/GPa) | $\gamma$ | $\omega_{exp}^{b}$ (cm$^{-1}$) | $\omega_{exp}^{c}$ (cm$^{-1}$) | $d\omega/dP_{exp}^{c}$ (cm$^{-1}$/GPa) |
|---|---|---|---|---|---|---|---|---|---|---|
| $B_g$ | 70 | -0.46 | -0.69 | 66 | 64 | | | | | |
| $A_g$ | 72 | 0.42 | 0.61 | 65 | 61 | | | | | |
| $A_g$ | 91 | 0.04 | 0.05 | 84 | 88 | -0.1 | -0.12 | | | |
| $B_g$ | 92 | -0.75 | -0.86 | 91 | 102 | | | | | |
| $A_g$ | 102 | 0.17 | 0.18 | 97 | 102 | 0.2 | 0.21 | | | |
| $B_g$ | 115 | 1.60 | 1.46 | 108 | 115 | | | | | |
| $B_g$ | 127 | 1.02 | 0.84 | 120 | 127 | | | 127 | 124.2 | |
| $A_g$ | 134 | 0.15 | 0.12 | 126 | 137 | 0.0 | 0.0 | | 138.4 | |
| $A_g$ | 143 | 1.66 | 1.22 | 136 | 146 | 1.4 | 1.00 | 147 | 143.8 | |
| $A_g$ | 154 | 1.80 | 1.23 | 142 | 160 | 2.3 | 1.51 | 158 | 156.6 | |
| $B_g$ | 158 | 2.17 | 1.44 | 143 | 158 | | | | | |
| $B_g$ | 183 | 2.80 | 1.61 | 170 | 189 | | | 189 | 187.3 | |
| $A_g$ | 188 | 3.13 | 1.75 | 173 | 193 | | | | | |
| $B_g$ | 204 | 2.60 | 1.34 | 178 | 209 | | | 208 | 204.7 | |
| $B_g$ | 224 | 3.40 | 1.59 | 210 | 232 | | | | | |
| $A_g$ | 230 | 2.70 | 1.23 | 203 | 235 | 3.1 | 1.38 | 238 | 242.1 | |
| $B_g$ | 242 | 3.36 | 1.46 | 218 | 241 | | | | | |
| $A_g$ | 252 | 3.28 | 1.37 | 232 | 252 | 3.7 | 1.54 | 251 | 260.4 | |
| $B_g$ | 297 | 0.90 | 0.32 | 292 | 309 | | | 309 | 306.2 | 0.39 |
| $B_g$ | 316 | 2.14 | 0.71 | 315 | 331 | | | | | |
| $A_g$ | 317 | 0.70 | 0.23 | 310 | 326 | 0.8 | 0.26 | 329 | 326.3 | 2.60 |
| $A_g$ | 336 | 2.41 | 0.75 | 324 | 349 | 2.5 | 0.75 | 349 | 345.8 | 2.97 |
| $A_g$ | 355 | 3.19 | 0.94 | 334 | 373 | 2.8 | 0.79 | 374 | 370.7 | 3.15 |
| $A_g$ | 380 | 3.51 | 0.97 | 367 | 397 | 3.2 | 0.85 | 398 | 394.8 | 3.73 |
| $B_g$ | 389 | 2.44 | 0.66 | 378 | 400 | | | | | |
| $B_g$ | 410 | 2.10 | 0.54 | 394 | 426 | | | | 420.7 | |
| $A_g$ | 423 | 2.28 | 0.57 | 405 | 439 | 2.0 | 0.48 | 440 | 436.4 | 2.44 |
| $B_g$ | 427 | 3.80 | 0.93 | 406 | 440 | | | | 463.4 | |
| $A_g$ | 784 | 3.83 | 0.51 | 865 | 768 | 4.1 | 0.56 | 770 | 766.5 | 4.50 |
| $B_g$ | 799 | 3.93 | 0.52 | 877 | 790 | | | | 782.9 | |
| $A_g$ | 806 | 4.49 | 0.58 | 883 | 794 | 4.9 | 0.65 | 794 | 792.1 | 3.73 |
| $A_g$ | 836 | 2.60 | 0.33 | 916 | 819 | 2.5 | 0.32 | 819 | 817.5 | 2.30 |
| $B_g$ | 850 | 4.78 | 0.59 | 911 | 843 | | | | 826.0 | |
| $B_g$ | 861 | 3.93 | 0.48 | 934 | 855 | | | | 840.9 | 3.90 |
| $A_g$ | 870 | 3.23 | 0.39 | 925 | 859 | 3.6 | 0.44 | 859 | 856.5 | 4.05 |
| $B_g$ | 892 | 2.21 | 0.26 | 965 | 882 | 4.8 | 0.57 | | | |

**Table 3:** Raman frequencies (ω) and pressure coefficients (dω/dP) for the high-pressure phase determined from theory (16.4 GPa) and experiments (15.4 GPa). The Grüneisen parameter was calculated according to $\gamma = \frac{B_0}{\omega_0}\frac{\partial \omega}{\partial P}$ using $B_0 = 154$ GPa.

| Raman mode | $\omega_{theo}$ (cm$^{-1}$) | $d\omega/dP_{theo}$ (cm$^{-1}$/GPa) | γ | $\omega_{exp}$ (cm$^{-1}$) | $d\omega/dP_{exp}$ (cm$^{-1}$/GPa) |
|---|---|---|---|---|---|
| Ag | 75.90 | 0.33 | 0.70 | | |
| Bg | 80.90 | 0.82 | 1.67 | | |
| Bg | 83.25 | - 0.72 | -1.24 | | |
| Ag | 95.80 | 0.09 | 0.14 | | |
| Bg | 104.10 | 0.32 | 0.49 | | |
| Ag | 108.68 | 0.56 | 0.85 | 114 | 1.3 |
| Ag | 133.62 | 1.47 | 1.95 | 129 | 1.5 |
| Bg | 135.58 | 1.68 | 2.17 | | |
| Ag | 139.47 | 1.86 | 2.45 | | |
| Bg | 138.17 | 0.71 | 0.86 | | |
| Ag | 152.78 | 0.90 | 0.98 | 153 | 2.2 |
| Bg | 157.82 | 2.49 | 2.93 | 159 | 3.0 |
| Ag | 162.91 | 1.32 | 1.39 | | |
| Bg | 174.10 | 1.32 | 1.30 | | |
| Ag | 183.99 | 2.12 | 2.09 | 181 | 3.4 |
| Bg | 194.54 | 2.47 | 2.34 | 197 | 4.1 |
| Bg | 212.27 | 3.28 | 2.88 | | |
| Ag | 217.73 | 3.22 | 2.75 | | |
| Bg | 220.36 | 0.05 | 0.03 | | |
| Ag | 231.13 | 3.42 | 2.75 | 227 | 3.2 |
| Bg | 235.87 | 3.36 | 2.61 | | |
| Ag | 242.14 | 3.75 | 2.95 | | |
| Ag | 254.50 | 3.71 | 2.77 | | |
| Bg | 259.40 | 5.33 | 4.18 | 258 | 4.9 |
| Bg | 265.35 | 2.99 | 2.07 | | |
| Ag | 268.39 | 3.35 | 2.30 | | |
| Ag | 276.84 | 3.02 | 1.99 | 277 | 2.9 |
| Bg | 276.32 | 2.49 | 1.59 | | |
| Bg | 279.59 | 3.12 | 2.05 | | |
| Ag | 289.49 | 4.46 | 2.95 | 286 | 3.3 |
| Ag | 305.20 | 2.54 | 1.47 | 302 | 4.1 |
| Ag | 317.00 | 2.13 | 1.13 | 318 | 5.0 |
| Ag | 329.92 | 4.54 | 2.57 | 327 | 6.2 |
| Ag | 338.66 | 2.75 | 1.41 | | |
| Bg | 341.75 | 3.28 | 1.67 | | |
| Ag | 351.43 | 3.24 | 1.63 | | |
| Bg | 357.70 | 2.82  | 1.34  | 367 | 2.8 |
| Bg | 378.18 | 4.97  | 2.37  | 386 | 4.3 |
| Bg | 380.42 | - 0.52| -0.21 |     |     |
| Ag | 386.02 | 2.98  | 1.32  |     |     |
| Ag | 395.95 | 2.32  | 0.97  |     |     |
| Bg | 399.08 | 2.61  | 1.10  | 400 | 3.3 |
| Bg | 411.49 | 3.34  | 1.38  | 417 | 4.1 |
| Ag | 411.59 | 1.33  | 0.52  |     |     |
| Ag | 430.57 | 2.06  | 0.78  |     |     |
| Bg | 433.72 | 2.34  | 0.89  | 432 | 5.1 |
| Bg | 435.28 | 0.16  | 0.05  |     |     |
| Ag | 437.34 | 1.84  | 0.69  |     |     |
| Ag | 450.54 | 3.12  | 1.18  | 454 | 2.6 |
| Ag | 472.10 | 2.16  | 0.75  |     |     |
| Bg | 480.41 | 3.09  | 1.07  | 473 | 3.6 |
| Bg | 488.66 | 2.30  | 0.78  | 489 | 4.4 |
| Ag | 504.275| 2.85  | 0.93  |     |     |
| Bg | 526.90 | 3.57  | 1.14  | 527 | 4.5 |
| Bg | 543.27 | 4.16  | 1.31  |     |     |
| Ag | 555.97 | 3.25  | 0.98  | 559 | 5.4 |
| Bg | 570.69 | 3.45  | 1.02  | 575 | 5.5 |
| Bg | 602.86 | 2.37  | 0.64  |     |     |
| Ag | 615.65 | 3.03  | 0.81  |     |     |
| Ag | 628.78 | 7.32  | 2.10  |     |     |
| Ag | 663.06 | 0.81  | 0.19  | 654 | 3.4 |
| Ag | 714.38 | 4.25  | 0.99  | 689 | 3.4 |
| Bg | 726.98 | 3.90  | 0.89  |     |     |
| Bg | 744.16 | 4.06  | 0.91  | 747 | 2.9 |
| Bg | 789.48 | 3.34  | 0.69  | 767 | 3.6 |
| Bg | 818.77 | 3.39  | 0.67  | 810 | 2.6 |
| Ag | 841.86 | 2.96  | 0.57  | 847 | 4.2 |
| Bg | 854.69 | 2.36  | 0.44  |     |     |
| Ag | 862.62 | 2.30  | 0.42  |     |     |
| Bg | 870.13 | 3.05  | 0.56  |     |     |
| Ag | 930.66 | 2.34  | 0.40  | 922 | 3.5 |
| Bg | 940.32 | 2.33  | 0.39  |     |     |

**Table 4:** Calculated structure of BaWO$_4$ II-type LaVO$_4$ at 16.4 GPa ($a$ = 12.319 Å, $b$ = 6.443 Å, $c$ = 6.775 Å, and $\beta$ = 95.80º). Wyckoff positions are indicated. To facilitate comparison we include here the unit-cell parameters determined from the experiments at 16.0(1) GPa: $a$ = 12.289(9) Å, $b$ = 6.492(5) Å, $c$ = 6.836(5) Å, and $\beta$ = 95.6(1)º.

| Atom | x | y | z |
|---|---|---|---|
| La$_1$ (4e) | 0.89138 | 0.34364 | 0.11856 |
| La$_2$ (4e) | 0.87725 | 0.05433 | 0.63987 |
| V$_1$ (4e) | 0.86896 | 0.83095 | 0.17888 |
| V$_2$ (4e) | 0.83191 | 0.55824 | 0.60671 |
| O$_1$ (4e) | 0.93820 | 0.04308 | 0.31838 |
| O$_2$ (4e) | 0.79212 | 0.35971 | 0.77788 |
| O$_3$ (4e) | 0.90611 | 0.40060 | 0.47845 |
| O$_4$ (4e) | 0.77509 | 0.63949 | 0.09695 |
| O$_5$ (4e) | 0.91628 | 0.69045 | 0.87024 |
| O$_6$ (4e) | 0.81457 | 0.77366 | 0.42413 |
| O$_7$ (4e) | 0.99659 | 0.70000 | 0.15006 |
| O$_8$ (4e) | 0.82884 | 0.01541 | 0.98855 |

**Table 5:** (0 GPa) Theoretical infrared frequencies (ω), pressure coefficients (dω/dP), and Grüneisen parameters γ, for the low-pressure phase.

| Infrared mode | ω (cm$^{-1}$) | dω/dP (cm$^{-1}$/GPa) | γ | Infrared mode | ω (cm$^{-1}$) | dω/dP (cm$^{-1}$/GPa) | γ |
|---|---|---|---|---|---|---|---|
| $A_u$ | 81 | 0.35 | 0.45 | $A_u$ | 290 | 0.39 | 0.14 |
| $B_u$ | 90 | 2.93 | 3.43 | $B_u$ | 342 | 4.45 | 1.37 |
| $A_u$ | 101 | 2.29 | 2.39 | $A_u$ | 355 | 3.43 | 1.02 |
| $A_u$ | 118 | 0.38 | 0.34 | $A_u$ | 372 | 3.57 | 1.01 |
| $B_u$ | 128 | -0.96 | -0.79 | $B_u$ | 378 | 1.70 | 0.47 |
| $A_u$ | 150 | 3.03 | 2.13 | $B_u$ | 416 | 2.53 | 0.64 |
| $B_u$ | 154 | 0.62 | 0.42 | $B_u$ | 419 | 3.72 | 0.93 |
| $B_u$ | 164 | 4.08 | 2.62 | $A_u$ | 466 | 3.02 | 0.68 |
| $A_u$ | 165 | 2.36 | 1.51 | $A_u$ | 782 | 3.94 | 0.53 |
| $B_u$ | 171 | 4.27 | 2.63 | $B_u$ | 790 | 3.33 | 0.44 |
| $A_u$ | 205 | 3.73 | 1.92 | $B_u$ | 806 | 4.46 | 0.58 |
| $B_u$ | 221 | 3.72 | 1.77 | $A_u$ | 821 | 5.15 | 0.66 |
| $A_u$ | 223 | 1.03 | 0.48 | $B_u$ | 842 | 5.33 | 0.67 |
| $A_u$ | 246 | 3.53 | 1.51 | $A_u$ | 843 | 4.49 | 0.56 |
| $B_u$ | 252 | 1.93 | 0.81 | $B_u$ | 854 | 3.98 | 0.49 |
| $A_u$ | 279 | 2.56 | 0.97 | $A_u$ | 878 | 3.90 | 0.46 |
| $B_u$ | 290 | 2.24 | 0.81 | | | | |

**Table 6:** (16.4 GPa) Theoretical infrared frequencies (ω), pressure coefficients (dω/dP), and Grüneisen parameters γ, for the high-pressure phase.

| Raman mode | ω (cm$^{-1}$) | dω/dP (cm$^{-1}$/GPa) | γ | Raman mode | ω (cm$^{-1}$) | dω/dP (cm$^{-1}$/GPa) | Raman mode |
|---|---|---|---|---|---|---|---|
| A$_u$ | 63.36  | - 0.58 | -1.31 | B$_u$ | 375.48 | 4.98   | 2.38  |
| B$_u$ | 72.00  | - 0.52 | -1.03 | B$_u$ | 376.99 | - 0.11 | -0.04 |
| B$_u$ | 99.15  | 0.46   | 0.75  | A$_u$ | 392.93 | 1.89   | 0.80  |
| A$_u$ | 118.76 | 1.67   | 2.56  | B$_u$ | 397.81 | 3.42   | 1.49  |
| A$_u$ | 128.66 | 1.08   | 1.45  | B$_u$ | 400.16 | 2.34   | 0.99  |
| B$_u$ | 134.27 | 0.64   | 0.77  | A$_u$ | 426.03 | 2.81   | 1.10  |
| B$_u$ | 141.24 | 1.45   | 1.81  | B$_u$ | 435.46 | 2.11   | 0.79  |
| A$_u$ | 142.19 | 1.58   | 1.97  | A$_u$ | 436.00 | 2.62   | 0.99  |
| A$_u$ | 165.66 | 3.10   | 3.60  | A$_u$ | 445.36 | 1.81   | 0.67  |
| A$_u$ | 166.75 | 0.31   | 0.30  | B$_u$ | 454.15 | 2.19   | 0.79  |
| B$_u$ | 175.25 | 2.05   | 2.10  | A$_u$ | 455.12 | 2.24   | 0.81  |
| A$_u$ | 179.88 | 1.02   | 0.96  | B$_u$ | 463.45 | 2.55   | 0.92  |
| B$_u$ | 191.21 | 1.97   | 1.84  | A$_u$ | 482.96 | 2.89   | 1.01  |
| A$_u$ | 212.11 | 2.84   | 2.38  | B$_u$ | 487.81 | 1.87   | 0.62  |
| B$_u$ | 213.72 | 2.39   | 1.98  | A$_u$ | 502.11 | 3.04   | 1.01  |
| B$_u$ | 222.79 | 3.63   | 3.11  | B$_u$ | 529.77 | 3.58   | 1.13  |
| A$_u$ | 227.83 | 2.88   | 2.30  | B$_u$ | 540.93 | 3.87   | 1.22  |
| B$_u$ | 231.49 | 2.87   | 2.23  | A$_u$ | 542.91 | 2.91   | 0.89  |
| A$_u$ | 234.93 | 3.33   | 2.67  | B$_u$ | 565.89 | 3.46   | 1.03  |
| B$_u$ | 234.93 | 1.11   | 0.79  | B$_u$ | 598.62 | 2.82   | 0.78  |
| A$_u$ | 250.69 | 4.63   | 3.61  | A$_u$ | 612.54 | 4.07   | 1.12  |
| A$_u$ | 254.83 | 0.95   | 0.62  | A$_u$ | 629.77 | 6.56   | 1.86  |
| B$_u$ | 265.45 | 4.25   | 3.08  | A$_u$ | 663.42 | 0.93   | 0.22  |
| B$_u$ | 272.63 | 2.84   | 1.89  | B$_u$ | 698.85 | 4.39   | 1.05  |
| A$_u$ | 278.73 | 3.26   | 2.11  | A$_u$ | 712.96 | 4.56   | 1.07  |
| A$_u$ | 294.95 | 4.12   | 2.61  | B$_u$ | 738.07 | 4.24   | 0.96  |
| B$_u$ | 299.34 | - 0.22 | 2.31  | B$_u$ | 793.26 | 3.37   | 0.69  |
| A$_u$ | 299.70 | 2.71   | 1.59  | A$_u$ | 810.51 | 3.08   | 0.61  |
| B$_u$ | 318.25 | 3.27   | 1.84  | A$_u$ | 841.29 | 2.29   | 0.43  |
| A$_u$ | 324.15 | 5.86   | 3.43  | B$_u$ | 850.11 | 2.49   | 0.47  |
| A$_u$ | 328.21 | - 0.54 | -0.26 | B$_u$ | 873.24 | 2.67   | 0.49  |
| B$_u$ | 338.31 | 3.65   | 1.92  | A$_u$ | 898.44 | 2.56   | 0.45  |
| A$_u$ | 343.47 | 2.54   | 1.26  | B$_u$ | 929.30 | - 0.22 | 0.37  |
| B$_u$ | 348.40 | 2.61   | 1.27  | A$_u$ | 929.98 | 2.23   | 0.38  |
| A$_u$ | 369.87 | 3.12   | 1.44  |       |        |        |       |

**Figure Captions**

**Figure 1: (color online)** Schematic structure of the crystal structure of the low-pressure (left) and high-pressure phase of LaVO$_4$. The coordination polyhedra of V and La are shown in red and green, respectively.

**Figure 2: (color online)** Unit-cell parameters and volume versus pressure. Solid squares and circles are from Refs. 18 and 9, respectively. Empty circles and diamonds represent the results obtained in this work from PXRD measurements for the low-pressure and high-pressure phase, respectively. Empty squares show the results obtained in this work from SXRD measurements for the low-pressure. The dashed lines represent the theoretical results. Solid lines are just a guide to the eye. For the HP phase we plotted *a*/2 instead of *a*, and *V*/2 instead of *V* for the sake of comparison.

**Figure 3:** Ambient pressure Raman spectra for different polarizations together with mode positions indicated by ticks. See the text for a description of the different polarizations.

**Figure 4:** Pressure-dependent sequence of selected high-pressure Raman spectra. "r" indicates sample recovered after decompression.

**Figu**re 5: Pressure evolution of Raman modes. Circles: low-pressure phase. Squares: high-pressure phase. Solid (empty) symbols correspond to pressure increase (release). The lines show linear fits to the experimental data. The vertical lines labeled as upstroke (downstroke) indicated the lowest pressure where the high-pressure phase is identified under compression (decompression).

**Figure 6:** Powder XRD patterns measured for the low-pressure phase at 11.5 GPa and the high-pressure phase at 16 GPa. Symbols: experiments. Solid lines: Rietveld refinements and residuals. Ticks indicate the positions of Bragg reflections. Ne peaks are indicated.

**Figure 7:** Calculated enthalpy difference per formula unit versus pressure.

**Figure 1**

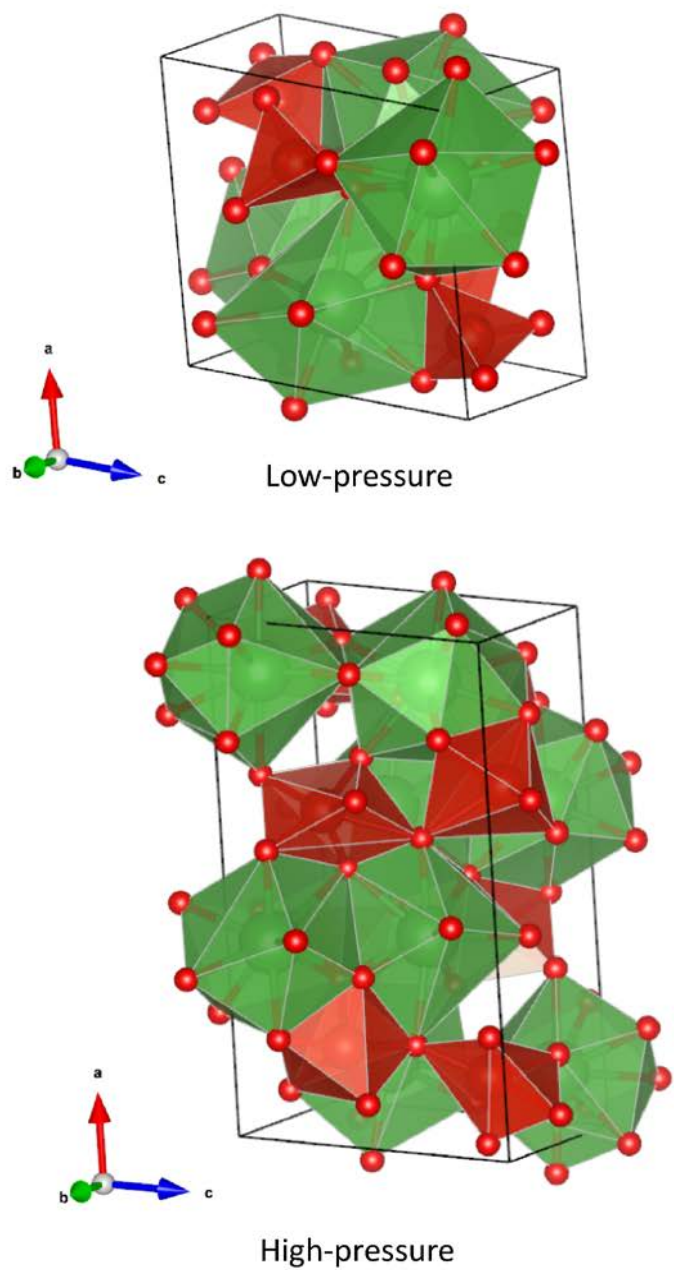

Low-pressure

High-pressure

**Figure 2**

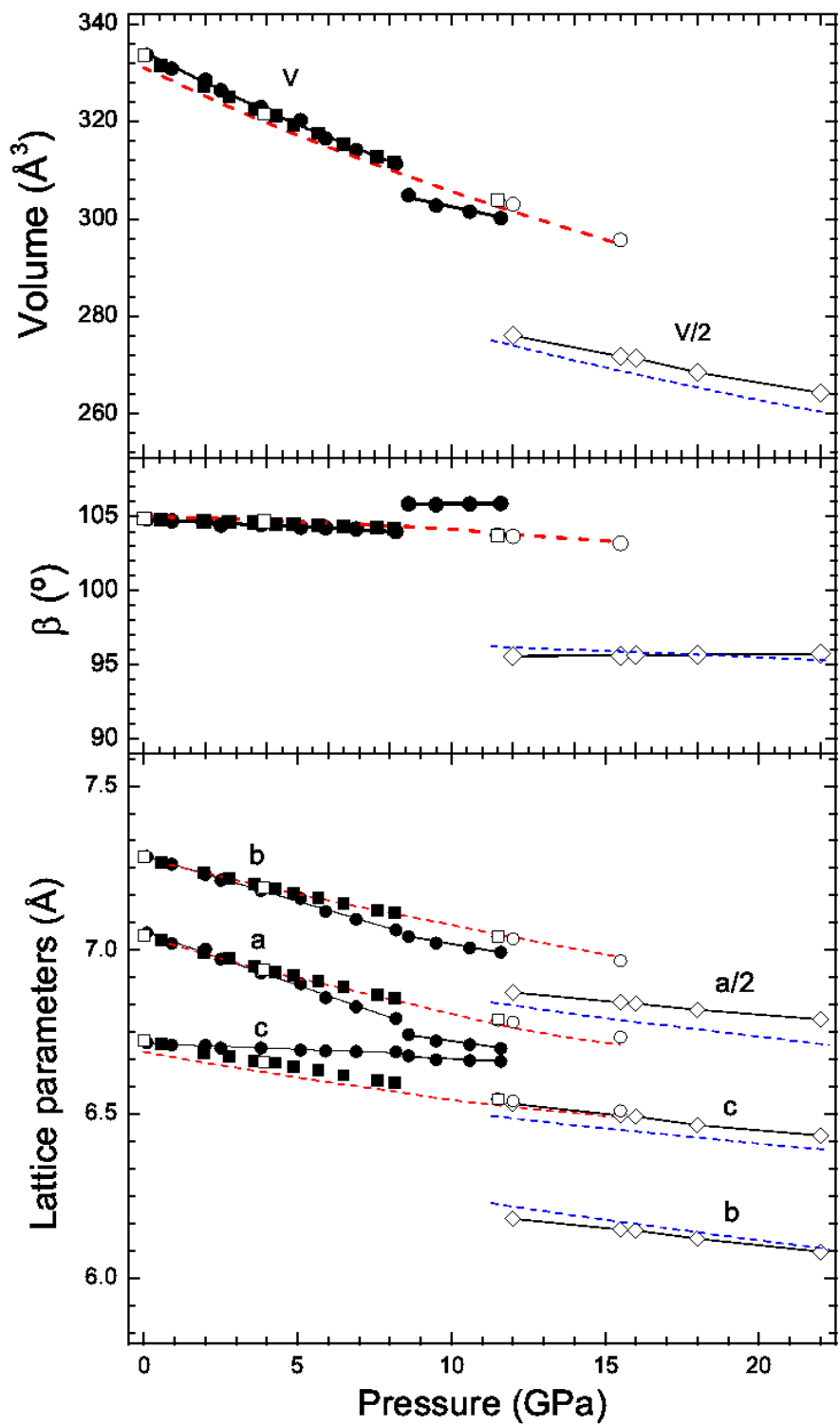

**Figure 3**

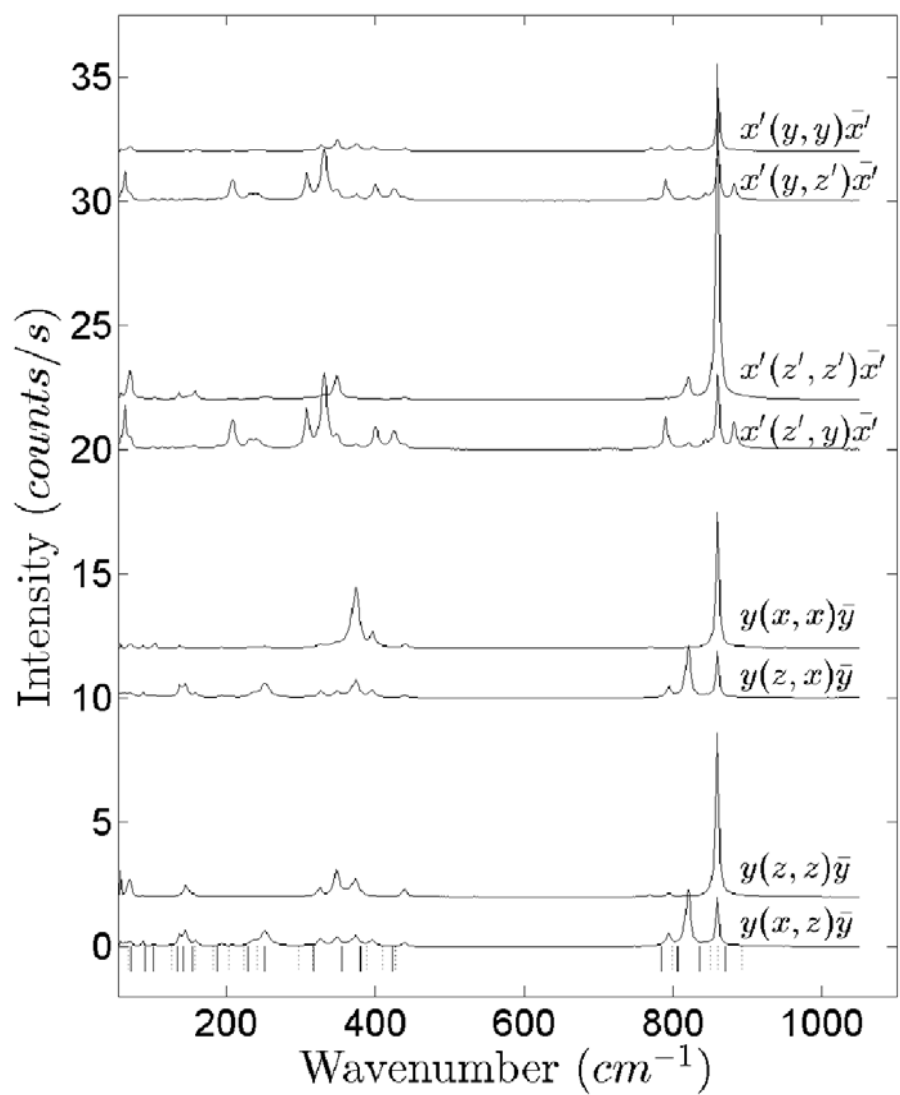

**Figure 4**

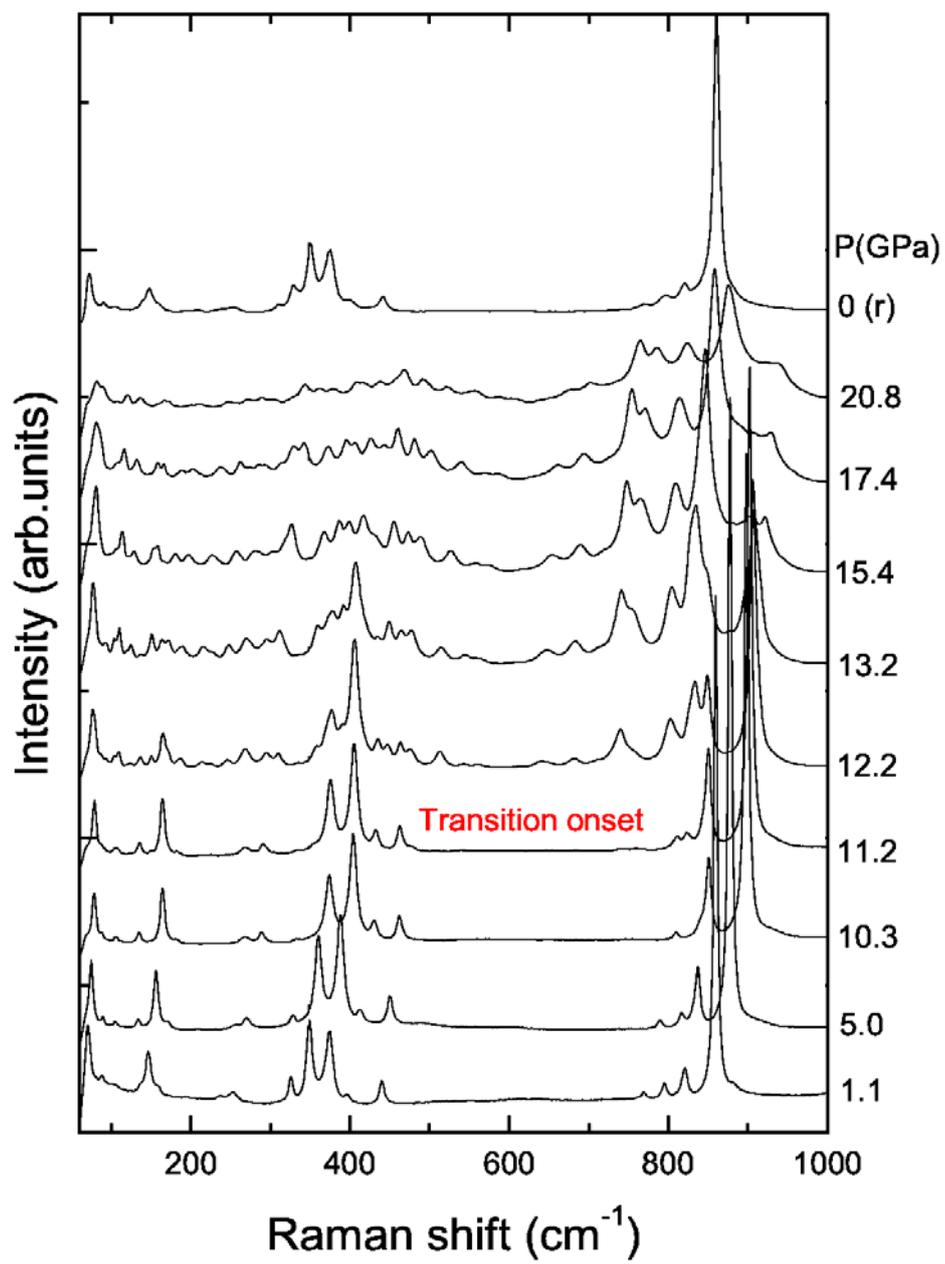

**Figure 5**

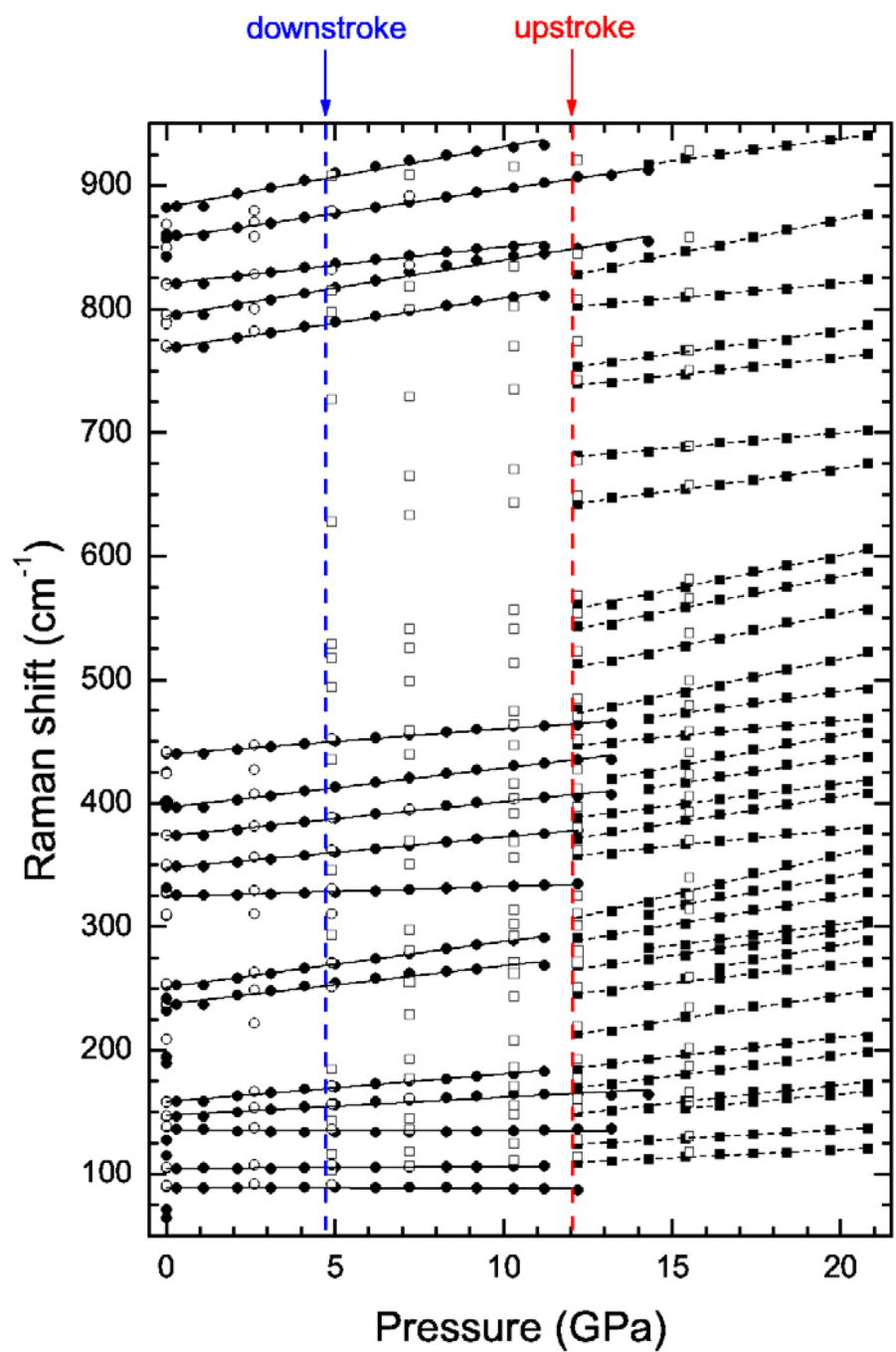

**Figure 6**

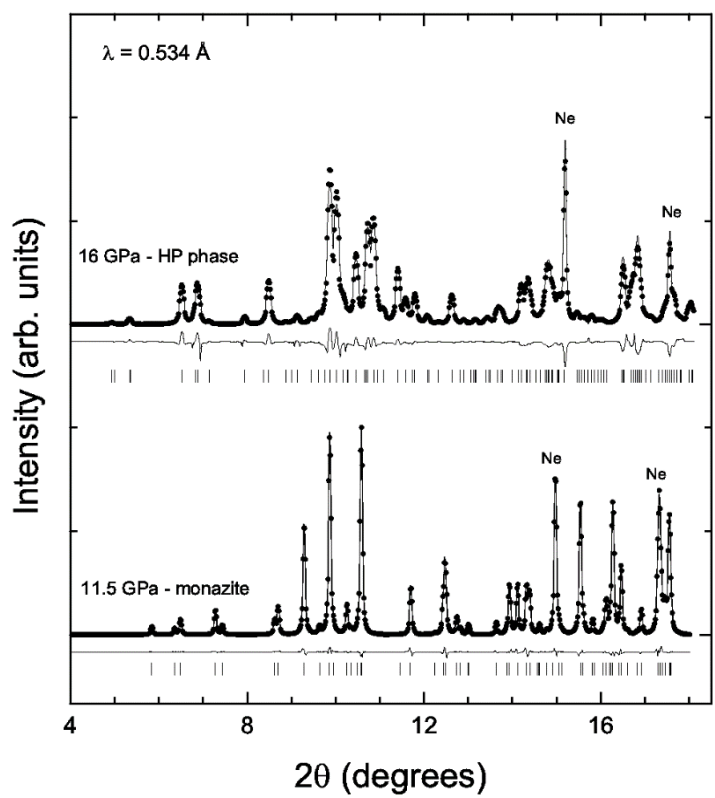

**Figure 7**

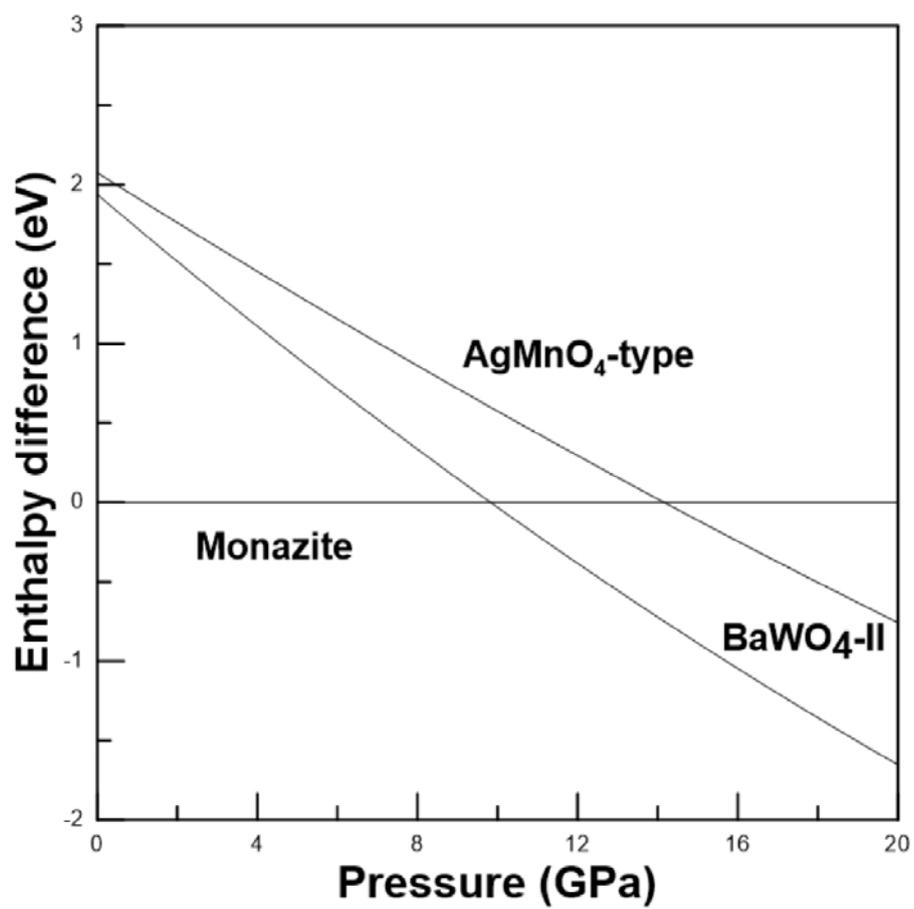